\documentclass[aps,prb,twocolumn,showpacs,amsmath,amssymb,superscriptaddress]{revtex4-2}
\usepackage[utf8]{inputenc}
\usepackage{amsmath}
\usepackage{graphicx}
\usepackage{hyperref}
\usepackage{bbold}
\usepackage{braket}
\usepackage{etoolbox}
\usepackage{cancel}
\usepackage{mathtools}
\usepackage{color}

\newcommand{\vb}{|\hspace{-0.9mm}}

\begin{document}

\title{Absence of thermalization of free systems coupled to gapped interacting reservoirs} 
    \author{Marko Ljubotina}
    \affiliation{Physics Department, Faculty of Mathematics and Physics, University of Ljubljana, 1000 Ljubljana, Slovenia}
    \affiliation{IST Austria, Am Campus 1, 3400 Klosterneuburg, Austria}
    \author{Dibyendu Roy}
    \affiliation{Raman Research Institute, Bangalore 560080, India}
    \author{Tomaž Prosen}
    \affiliation{Physics Department, Faculty of Mathematics and Physics, University of Ljubljana, 1000 Ljubljana, Slovenia}
    
    \begin{abstract}
        We study the thermalization of a small $XX$ chain coupled to long, gapped $XXZ$ leads at either side by observing the relaxation dynamics of the whole system. 
        Using extensive tensor network simulations, we show that such systems, although not integrable, appear to show either extremely slow thermalization or even lack thereof since the two can not be distinguished within the accuracy of our numerics. 
        We show that the persistent oscillations observed in the spin current in the middle of the $XX$ chain are related to eigenstates of the entire system located within the gap of the boundary chains. 
        We find from exact diagonalization that some of these states remain strictly localized within the $XX$ chain and do not hybridize with the rest of the system. 
        The frequencies of the persistent oscillations determined by numerical simulations of  dynamics match the energy differences between these states exactly.
        This has important implications for open systems, where the strongly interacting leads are often assumed to thermalize the central system. 
        Our results suggest that if we employ gapped systems for the leads, this assumption does not hold; this finding is particularly relevant to any potential future experimental studies of open quantum systems.
    \end{abstract}
    \maketitle

    \textbf{Introduction.~} 
    Microscopic analysis of quantum transport, especially nonequilibrium transport across a system, is often carried out by connecting the boundaries of the system to two or more reservoirs \cite{BenentiReview,PolettiReview,RMP}. 
    These reservoirs act as a source or sink of conserved quantities, e.g., charge, energy, spin. 
    They are generally maintained at a fixed bias by keeping them, e.g., at different chemical potentials or temperatures or magnetizations. 
    The explicit inclusion of such interface between a system and a reservoir is essential in understanding the role of decoherence and dissipation (induced by the reservoir to the system) in controlling the quantum transport in physical systems \cite{Landauer57, Landauer70, Buttiker86, Kohler2005, nazarov_blanter_2009}. 
    For example, the difference between the Landauer's four-probe and two-probe resistance arises from the contact resistance due to such interfaces \cite{Landauer57,Landauer70,Picciotto01}. 
    Apart from acting as a perfect source or sink, a reservoir should also thermalize the system to which it is connected. 
    Therefore, it is crucial to investigate the essential properties of a reservoir to implement the above-prescribed tasks.
    
    In most theoretical studies, the reservoirs are traditionally modeled as large (infinite) collections of harmonic oscillators or noninteracting electrons or spins \cite{Weiss12, Kohler2005, DharPRB2003, Segal2003, DharRoy2006, Karevski09, Arrachea09}.
    The large reservoir volume limit is taken to ensure very long (practically infinite) Poincare recurrence time so that the 
    particle, or any local packet of a conserved quantity, once transferred to the reservoir is not fed back to the system. 
    Nevertheless, these noninteracting models of reservoirs neither thermalize themselves in strict quantum mechanical sense nor do they provide any mechanism for mixing or relaxation for the transported degrees of freedom entering from the system. 
    One then assumes some thermal statistical ensemble (micro-canonical or canonical or grand-canonical) for these reservoirs from the outset. 
    However, the absence of mixing in the noninteracting reservoirs creates further issues when their energy dispersion does not cover all the system's energy levels. 
    For example, it has already been demonstrated that a system is not thermalized by such noninteracting reservoirs when one or more energy levels (bound states) of it lie either outside the reservoir's energy band or inside the energy gap of the reservoir spectrum~\cite{DharPRB2006, Zgirski11,Souto16, Bondyopadhaya_2019}. 
    These bound states are mostly localized within the central system's Hilbert space and do not leak out in the reservoir. 
    Notably, there is no unique long-time steady-state for such a coupled system-reservoir in the presence of bound states, and transport properties across the system depend on the initialization of the full system.
    
    It is thus intriguing to examine if the inclusion of interactions between the reservoir's degrees of freedom can help in thermalizing the system with the reservoir in the presence of bound states, i.e. the central system eigenenergies gapped away from the spectrum of the infinite reserviors. 
    Such energy-gap in the spectrum of a reservoir with nontrivial interactions between its constituents can naturally occur in many physical set-ups where the reservoirs are superconductors, correlated magnets, or other strongly correlated materials. 
    One would naively expect that system-reservoir interactions would 
    mix the bound states and eventually thermalize the system with the reservoir. 
    Nevertheless, it is a nontrivial problem to analytically study such coupled system-reservoir as the dynamical evolution of inhomogeneous interacting systems is a daunting task. 
    On the other hand, the numerical techniques to calculate the dynamics of such coupled system-reservoir models are mostly limited to relatively small sizes and time scales. 
    In this paper, we critically investigate the role of interactions in thermalizing bound states by applying large-scale numerics using the Time-Evolving Block Decimation (TEBD) algorithm~\cite{Vidal-03-1, Vidal-04-1, Schollwock-11-1}. 
    
\begin{figure}[ht!]
        \centering
        \includegraphics[width=0.95\columnwidth]{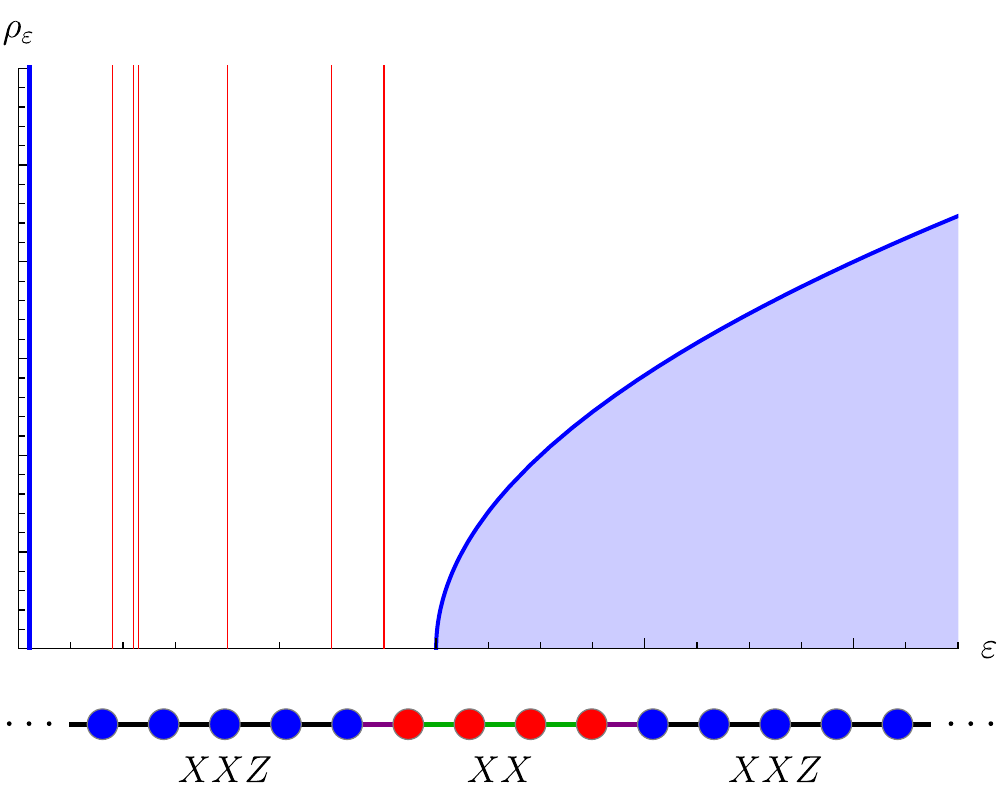}
        \caption{(Bottom) A schematic of a finite-size $XX$ spin-1/2 chain (red circles) sandwiched between two long anisotropic Heisenberg ($XXZ$) spin-1/2 chains (blue circles). (Top) Density of states $\rho_{\epsilon}$ with energy $\epsilon$ of the sandwich depicting discrete energy (bound) states (red lines) arising from $XX$ spin chain in the excitation gap between ground and first excited states of the boundary $XXZ$ spin chains.}
        \label{fig:schematic}
    \end{figure}

    We mainly study quench dynamics in a quantum $XX$ spin-1/2 chain coupled to two anisotropic Heisenberg ($XXZ$) spin-1/2 chains, one on either side (see Fig.~\ref{fig:schematic}), by following the time evolution of the full density matrix after the quench. 
    The $XXZ$ spin chains with an excitation energy gap between the ground and excited states act as boundary reservoirs with nontrivial (non-quadratic) interactions between spins. 
    The quantum $XX$ spin chain of short length as the central system in the middle has a gapless continuous spectrum with spins that can be considered trivially (quadratic) interacting. 
    We take very long lengths (up to 349 spins) for each boundary chains in our numerics to avoid Poincare recurrence and boundary effects. 
    The interface between the $XXZ$ and $XX$ chain is described by an $XX$ exchange coupling. 
    We choose the initial state of each boundary $XXZ$ chain in some equilibrium state at very high temperatures and at finite or zero magnetization. 
    For the middle $XX$ chain, we consider various initial states different from the boundary $XXZ$ chains. 
    We take the product state of these three
    subsystems to describe the initial state, and evolve the coupled system-reservoir by the full Hamiltonian, including the interfaces. 
    We then calculate the time-evolution of magnetization and spin current in the entire system. 
    We are specifically interested in detecting the relaxation dynamics of these quantities inside the middle $XX$ chain. 
    We find a strong signature of very slow (or absent) relaxation of some particular modes of the magnetization and spin current inside the $XX$ spin chain. 
    The frequencies of these special slow modes are related to the energies of states within the excitation gaps of the boundary $XXZ$ chains that remain localized to the middle $XX$ chain. 
    
    \textbf{The model.~}
    The Hamiltonian of the coupled spin-1/2 chains can be written in a general form as
    \begin{equation}
        \mathcal{H}=-\sum_{l=1}^{N-1}\sum_{\alpha}J^{\alpha}_{l}\sigma_l^{\alpha}\sigma_{l+1}^{\alpha},\label{hamXXZ}
    \end{equation}
    where $\sigma_l^{\alpha},\alpha \in \{x,y,z\}$ is the Pauli matrix at site $l$, and $J^{\alpha}_{l}$ is the exchange coupling between the $\alpha$ component of spins at sites $l$ and $l+1$.
    The choice of exchange couplings defines the system, boundary reservoirs and their interfaces. 
    We choose the boundary chains of length $L$ in the regions $l\in[1,L]$ and $l\in[L+M+1,N]$, and the middle chain of length $M$ is at $l\in[L+1,L+M]$. 
    We take the exchange couplings for the boundary chain as $J^{x}_l=J^{y}_l=J^{z}_l/\Delta=J_0$. 
    The middle $XX$ chain has $J^{x}_l=J^{y}_l=J_1$ and $J^{z}_l=0$. 
    The $XX$-type couplings at the interfaces are $J^{x}_l=J^{y}_l=J',J^{z}_l=0$ for $l\in\{L,L+M\}$. 
    The spin current operator is defined as $I_l=2J_l(\sigma_l^y \sigma_{l+1}^x-\sigma_l^x \sigma_{l+1}^y)$, were $J_l$ is the value of the hopping between sites $l$ and $l+1$. 
    The expectation value of the spin current is defined as $\langle I_l(t) \rangle= {\rm Tr}\{\rho(t)I_l\}$ where $\rho(t)$ is the density matrix of the full system at time $t$.
    
    We mostly investigate the quench dynamics of coupled spin chains in the $XXZ$-$XX$-$XXZ$ configuration, where we choose $J_1/J_0=0.2$, $J'/J_0=0.05$ and $\Delta=2$. 
    The choice of parameters here is such that some energy states of the full system appear within the gap of the two boundary chains (which have a nonzero gap for $\Delta>1$) due to the coupling to the $XX$ chain. Additionally, we have also simulated the quench dynamics of coupled spin chains in the $XXX$-$XX$-$XXX$ configuration for comparison, since the $XXX$ boundary chains are gapless. In the latter case, we simply fix $\Delta=1$ leaving other parameters unchanged. 
    
    \textbf{Quench protocols and results.~}
    In our numerical investigation of quench dynamics, we typically set the initial density matrix of the two boundary $XXZ$ or $XXX$ chains to a high temperature state with constant magnetization in the $z$ direction.
    For the middle $XX$ chain, we choose either a flat or a random magnetization profile, where in case of a flat profile we choose a value different to that of the boundary chains. 
    In general the initial density matrix takes the product form
    \begin{equation}
       \rho(0)\propto\bigotimes_{l=1}^{N}e^{\mu_l\sigma^z_l}\,,
    \end{equation}
    where the parameters $\mu_l$ determine the initial magnetization at each site. 
    We choose $\mu_l$ to be small for all $l$, such that our state is effectively a high temperature state. 
    We then evolve the density matrix $\rho(t)$ of the full system from the initial density matrix $\rho(0)$ using $\rho(t)=e^{-i\mathcal{H}t/\hbar}\rho(0)e^{i\mathcal{H}t/\hbar}$, and follow the time evolution of relevant local observables, such as the spin and spin current densities. 
    
    \begin{figure*}[ht!]
        \centering
        \includegraphics[width=0.99\textwidth]{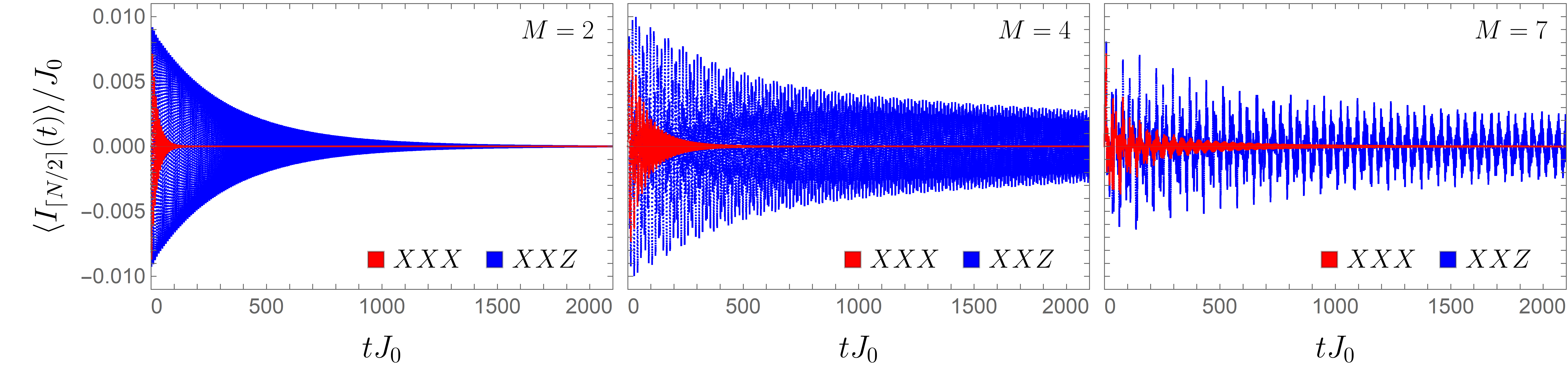}
        \caption{
           Time-evolution of spin current in the middle of $XX$ chain after the quench for both $XXZ$ (blue) and $XXX$ (red) boundary chains with various lengths of the middle segment ($M=2$, $M=4$ and $M=7$ from left to right). 
            A clear difference in relaxation dynamics is observed since the system with gapped boundaries exhibits much slower (or no) decay in spin current oscillations compared to the gapless boundaries. In the initial state, the boundaries are at infinite temperature ($\mu=0$) while the middle $XX$ chain has a random magnetisation profile ($\mu_l$ i.i.d. in $[-0.02,0.02]$). We use $J_1/J_0=0.2$ and $J'/J_0=0.05$ in all cases, as well as $\Delta=1$ for $XXX$ and $\Delta=2$ for $XXZ$. For all simulations, the length of the full system is set to $N=720$ with the $XX$ chain positioned in the middle. }
        \label{fig:compare}
    \end{figure*}
    
\begin{figure*}[ht!]
        \centering
        \includegraphics[width=0.99\textwidth]{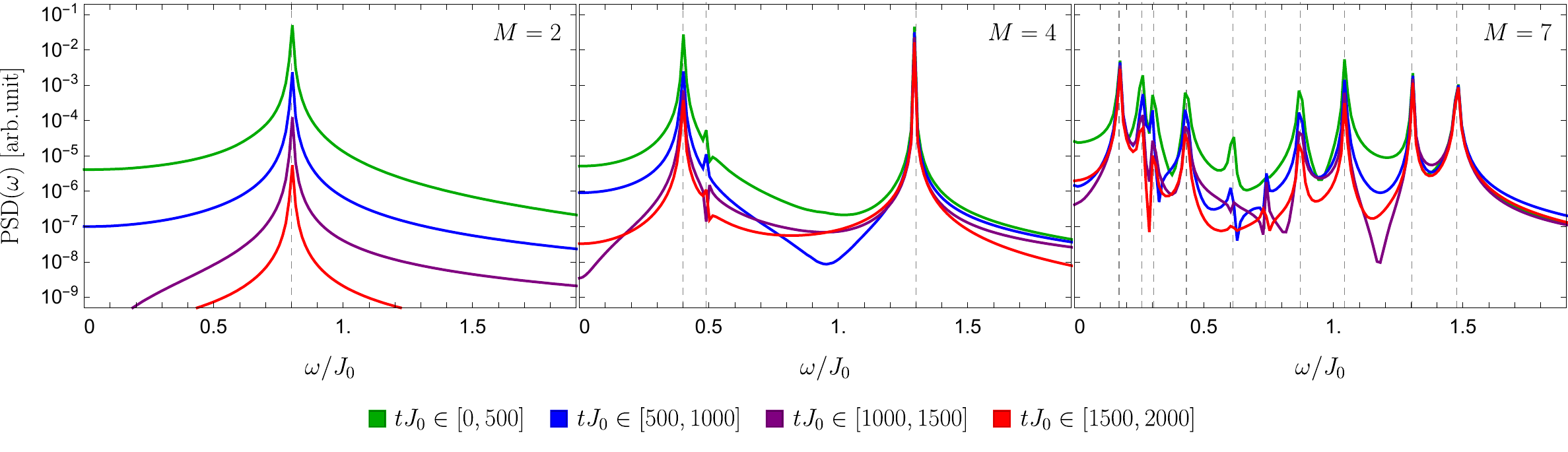}
        \caption{
            Power spectral density (PSD) extracted from different time windows (full coloured lines) of the TEBD simulation for several sizes $M$ of the middle chain of $XXZ$-$XX$-$XXZ$. 
            The gray dashed lines represent the frequencies of spin current oscillations obtained from bound state analysis using exact diagonalization on smaller systems. 
            We can observe excellent agreement between the peaks in PSD and the computed frequencies. 
            We see clear decay of the only existing frequency for $M=2$. However, as we increase $M$, we begin to see modes with extremely slow or even no decay within the accuracy of our simulations.
            This suggests that these modes would not thermalize with the rest of the system, and maintain a memory of the initial conditions for long times. 
        } 
        \label{fig:fourier}
    \end{figure*}

    In Fig.~\ref{fig:compare}, we compare the relaxation dynamics of $XXX$-$XX$-$XXX$ and $XXZ$-$XX$-$XXZ$ systems by looking at the time evolution of the spin current in the middle of the $XX$ chain. 
    It is immediately clear that in the case of $XXX$ boundary chains the system quickly relaxes as the middle chain thermalizes with the rest of the system. 
    We note here that even though the isolated $XXX$ and $XXZ$ chains are exactly solvable via Bethe ansatz, the coupled chains of $XXX$-$XX$-$XXX$ or $XXZ$-$XX$-$XXZ$ are not integrable \cite{Brenes18}. 
    Thus, the thermalization in $XXX$-$XX$-$XXX$ is in line with the eigenstate thermalization hypothesis (ETH)~\cite{ETHreview}. 
    On the other hand, looking at the system with $XXZ$ boundary chains we notice clearly slower themalization or even lack thereof at larger $M$. 
    Thus, the persistent oscillations of spin current in some particular frequency modes for larger $M$ seem to suggest a violation of ETH for $XXZ$-$XX$-$XXZ$ systems. 
    In Fig.~\ref{fig:fourier}, we further show the time dependence of the frequencies of these spin current oscillations in $XXZ$-$XX$-$XXZ$ systems for different $M$. 
    We observe that some frequencies are particularly stable and show (almost) no decay for greater $M$ within the accessible times. 
    We attribute this lack of thermalization in the $XXZ$ case to the existence of states due to the middle chain that are within the gap of the boundary chains. 
    To test this assertion, we now look at the eigenstate properties of shorter lengths of coupled chains paying particular attention to eigenstates within the spectral gap of the boundary chains. 
    
    \textbf{Eigenstate properties of the coupled chains.~} 
    
\begin{figure}[ht!]
        \includegraphics[width=0.99\linewidth]{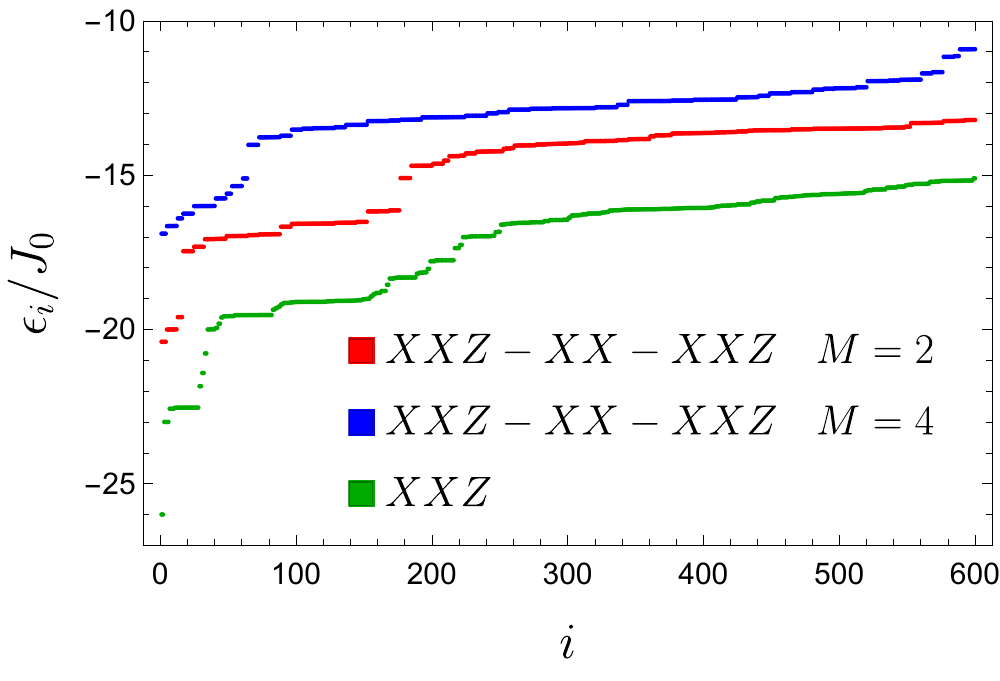}
        \caption{
            Typical low energy spectrum $\epsilon_i$ of an isolated $XXZ$ spin-1/2 chain (green) with an excitation gap, and coupled spin-1/2 chains of $XXZ$-$XX$-$XXZ$ configuration (red and blue) with bound states within the excitation gap. 
            The parameters are $J_1/J_0=0.2, J'/J_0=0.05$ and $\Delta=2$ in all plots. 
            The lengths are $L=N=14$ (green), $N=14, L=6, M=2$ (red), and $N=14, L=5, M=4$ (blue). 
        }
        \label{fig:excitation}
    \end{figure}

    In Fig.~\ref{fig:excitation}, we present a typical low energy spectrum of an isolated ferromagnetic $XXZ$ chain of length $N=14$ and $\Delta/J_0=2$. 
    It depicts an excitation gap $\delta E_{eg}/J_0=3.0$ between two degenerate ground states ($\epsilon_1/J_0=-26$) and the first excited states ($\epsilon_3/J_0=-23$). 
    In Fig.~\ref{fig:excitation}, we further show the energy eigenstates appearing within the excitation gap of isolated boundary chains in the the low energy spectrum of an $XXZ$-$XX$-$XXZ$ system for $N=14,L=6,M=2$ and $N=14,L=5,M=4$. 
    Increasing the length of the full system by lengthening the leads only shifts the values of these energies but does not change the differences between the energies of the bound states and thus has no effect on our results. 
   
    We notice that some of these mid-excitation-gap states have all spins pointing either up or down in both boundary chains and only have any non-trivial spin pattern in the middle chain. 
    In the presence of some local perturbation, we can thus have local spin dynamics between these special states which is confined in the middle chain without affecting the boundary chains. 
    Here we will refer to these states as bound states borrowing the notion from noninteracting models \cite{DharPRB2006,Bondyopadhaya_2019}. 
    For example, we have such special bound states with energies $\epsilon_i/J_0=-20.4,-19.6$ for $N=14,L=6,M=2$:
    \begin{equation}
        \begin{split}
            |-20.4\rangle_{\uparrow} &\approx \vb\uparrow \rangle_{L} \otimes (0.7068\vb\downarrow \uparrow \rangle+0.7068\vb\uparrow \downarrow \rangle) \otimes \vb\uparrow \rangle_{R}, \\
            |-19.6\rangle_{\uparrow} &\approx \vb\uparrow \rangle_{L} \otimes (-0.7066\vb\downarrow \uparrow \rangle+0.7066\vb\uparrow \downarrow \rangle) \otimes \vb\uparrow \rangle_{R}\,,
        \end{split}
        \nonumber
    \end{equation} 
    and $|-20.4\rangle_{\downarrow}$ and $ |-19.6\rangle_{\downarrow}$ with all spins flipped compared to $|-20.4\rangle_{\uparrow}$ and $ |-19.6\rangle_{\uparrow}$, respectively. 
    Here the basis states $\vb\downarrow\rangle_{L/R}$ and $\vb\uparrow\rangle_{L/R}$ represent the state of left or right boundary chain with all spins $\downarrow$ and $\uparrow$ respectively. 
    There are some other bound states of the form $\vb\uparrow \rangle_{L} \otimes \vb\uparrow \uparrow \rangle \otimes \vb\uparrow \rangle_{R}$, and $\vb\uparrow \rangle_{L} \otimes \vb\downarrow \downarrow \rangle \otimes \vb\uparrow \rangle_{R}$ as well as their spin-flipped partners at $\epsilon_i/J_0=-20$, which we do not consider here since they do not affect the spin current dynamics in the middle of $XX$ chain that we are interested in.     
    We emphasize that most eigenstates, including those with energy within the excitation gap of $XXZ$-$XX$-$XXZ$ system, are not of the above special forms, and have non-trivial spin texture both in the middle and boundary chains.  
    The bound states $|-20.4 \rangle_{\uparrow,\downarrow}$ and $|-19.6 \rangle_{\uparrow,\downarrow}$ still have some extremely small overlaps with the other basis states in their corresponding symmetry sectors. 
    Thus, the spin orientations of the boundary chains are mostly the same for the states $|-20.4\rangle_{\uparrow,\downarrow}$ and $|-19.6\rangle_{\uparrow,\downarrow}$. 
    A coherent oscillation of spin current at the middle of the $XX$ chain occurs at an angular frequency of $0.8J_0$ ($\hbar=1$) due to spin-exchange between $|-20.4\rangle_{\uparrow,\downarrow}$ and $|-19.6\rangle_{\uparrow,\downarrow}$ upon application of the current operator. 
    We find $\langle I_7(t) \rangle \propto \sin (0.8J_0t)$. 
    Nevertheless, the boundary chains directly connected to both of the spins in the middle act as a reservoir. 
    Thus, the boundary chains would induce strong dephasing/decoherence to such coherent oscillation, and the oscillations will eventually die out even in the presence of a large excitation gap. 
    While the coherent oscillation due to spin-flip inside the middle chain should also decay due to the dephasing effect from the boundary chains for $M=3$, it should survive for $M>3$ where a coherent spin-flip inside the middle chain between spins away from the interfaces is possible. 
    
    In order to observe such survival of oscillations, we extend the above analysis of bound states for an $XXZ$-$XX$-$XXZ$ system of $N=14, L=5, M=4$. 
    We find the energies of the bound states of our interest for this case are $\epsilon_i/J_0=-16.648,-16.249,-15.755,-15.354$, and the corresponding eigenstates are the following:
    \begin{equation}
        \begin{split}
            |-16.648\rangle_{\uparrow} &\approx \vb\uparrow \rangle_{L} \otimes (0.372\vb\downarrow \uparrow \uparrow \uparrow \rangle+ 0.601\vb\uparrow \downarrow \uparrow \uparrow \rangle\\&+0.601\vb\uparrow \uparrow \downarrow \uparrow \rangle + 0.372\vb\uparrow \uparrow \uparrow \downarrow\rangle) \otimes \vb\uparrow \rangle_{R}, \\
            |-16.249\rangle_{\uparrow} &\approx \vb\uparrow \rangle_{L} \otimes (-0.602\vb\downarrow \uparrow \uparrow \uparrow \rangle-0.371\vb\uparrow \downarrow \uparrow \uparrow \rangle\\&+0.371\vb\uparrow \uparrow \downarrow \uparrow \rangle + 0.602\vb\uparrow \uparrow \uparrow \downarrow\rangle) \otimes \vb\uparrow \rangle_{R}, \\
            |-15.755\rangle_{\uparrow} &\approx \vb\uparrow \rangle_{L} \otimes (0.601\vb\downarrow \uparrow \uparrow \uparrow \rangle-0.373\vb\uparrow \downarrow \uparrow \uparrow \rangle\\&-0.373\vb\uparrow \uparrow \downarrow \uparrow \rangle + 0.601\vb\uparrow \uparrow \uparrow \downarrow\rangle) \otimes \vb\uparrow \rangle_{R}, \\
            |-15.354\rangle_{\uparrow} &\approx \vb\uparrow \rangle_{L} \otimes (0.371\vb\downarrow \uparrow \uparrow \uparrow \rangle- 0.602\vb\uparrow \downarrow \uparrow \uparrow \rangle\\&+0.602\vb\uparrow \uparrow \downarrow \uparrow \rangle - 0.371\vb\uparrow \uparrow \uparrow \downarrow\rangle) \otimes \vb\uparrow \rangle_{R}\,,
        \end{split}
        \nonumber
    \end{equation} 
    and another four down eigenstates, $|-16.648\rangle_{\downarrow},|-16.249\rangle_{\downarrow}, |-15.755\rangle_{\downarrow}$ and $ |-15.354\rangle_{\downarrow}$, whose spins are all flipped with respect to the same energy up states given above. 
    Again, the bound states $|-16.648 \rangle_{\uparrow,\downarrow},|-16.249 \rangle_{\uparrow,\downarrow},|-15.755\rangle_{\uparrow,\downarrow}$ and $|-15.354\rangle_{\uparrow,\downarrow}$ have very small amplitudes to those basis states where one spin of the left or right boundary chain is flipped down. 
    The contribution of these bound states to the expectation value of the spin current at the middle of $XX$ chain is found to be
    \begin{equation}
        \begin{split}
        \hspace{-1.9mm}\langle I_7(t) \rangle &\propto (0.362c_0\sin(1.29J_0t)+0.223c_1\sin(0.4J_0t) \\&+0.138c_2\sin(0.49J_0t)-0.224c_3\sin(0.4J_0t))\,. \label{SC4}
        \end{split}
    \end{equation}
    
    Here the coefficients $c_0, c_1, c_2$, and $c_3$ depend on the initial state occupation amplitudes of these bound states, and can be considered real for simplicity. 
    The spin current in Eq.~\eqref{SC4} oscillates in time at angular frequencies $1.29J_0,0.4J_0$ and $0.49J_0$. 
    However, the amplitudes of these current oscillations essentially depend on the initialization (the coefficients $c_0,c_1,c_2,c_3$) of these bound states. 
    We further notice that the oscillation at angular frequency $1.29J_0$ is arising from the states $|-16.648\rangle_{\uparrow,\downarrow}$ and $|-15.354\rangle_{\uparrow,\downarrow}$ both of which have maximum amplitudes for spin flips of the middle spins (2nd and 3rd spin for $M=4$) away from the interface between the $XX$ chain and the boundary chains. 
    Any other frequency current oscillations do not have this feature. 
    Therefore, considering faster dephasing of those bound states with higher amplitudes for spin flips near the interfaces, we predict that the coherent oscillation at angular frequency $1.29J_0$ would survive for the longest time. 
    We can easily extend the above analysis to longer middle $XX$ chains.
    
    In Fig.~\ref{fig:fourier}, we now compare the frequencies of spin current oscillations obtained from the above analysis with bound states to the frequencies found from TEBD simulations. 
    The angular frequencies of current oscillations from TEBD match nicely with our predicted values from bound state analysis.
    Furthermore, we observe that our prediction on which modes will survive longest also holds, for instance, the oscillation at $\omega/J_0\approx1.29$ for $M=4$ shows very little decay compared to those at lower angular frequency. 
    A similar observation holds for $M=7$, where the oscillation at $\omega/J_0\approx1.476$ displays the least decay. 
    From this comparison, we can conclude that the extremely slow or even absence of thermalization indeed appears to be related to the bound states within the spectral gap of the boundary chains (acting as reservoirs), even in the presence of interactions within the boundary chains. 
    
    \textbf{Conclusion.~}
     In this work, we have checked the necessary conditions for a good thermalizing reservoir by extending the analysis from earlier works with noninteracting reservoirs \cite{DharPRB2006,Bondyopadhaya_2019} to interacting ones. 
    Using large-scale TEBD simulations, we showed that gapless interacting reservoirs appear to thermalize the system in question.
    On the other hand, interacting reservoirs with a spectral gap show prolonged, or even no thermalization \cite{Mossel_2010}. 
    This suggests the latter are not ideal candidates for coupling to systems to study dissipative dynamics of the central system.     
    Furthermore, we show that the frequencies of the persistent spin current oscillations we observed for $XXZ$ boundary chains match nicely to the the energy differences between the eigenstates that remain primarily localized within the middle chain and contribute to the spin current expectation value in the middle of the chain. 
    We have also investigated the role of stronger couplings between $XX$ and $XXZ$ chains, suggesting an inevitable change in relaxation dynamics as the mid-excitation-gap states shift with increasing couplings. 
    Our findings further indicate that an open quantum system description of transport depends not only on the properties of the system but also on those of the interfaces/couplings and reservoirs (e.g., spectral properties of reservoirs). Our study can be extended in understanding the relaxation dynamics of other correlated systems with various bound states, e.g., Yu-Shiba-Rusinov \cite{Yu65, Shiba68, Rusinov69} and Majorana \cite{Majorana37} bound states, which appear in the spectral gap of s-wave or p-wave superconductors. There are growing interests in understanding transport and nonequilibrium dynamics in inhomogeneous many-body quantum models by coupling multiple different systems \cite{Vasseur2014} especially quantum spin chains \cite{Biella19, delvecchio2021transport}, our present results would be helpful in those studies with a spectral gap that mainly was unexplored. 
    
    \textbf{Acknowledgements.}
    M.L. and T.P. acknowledge support from the European Research Council (ERC) through the advanced grant 694544 – OMNES and the grant P1-0402 of Slovenian Research Agency (ARRS).
    M.L. acknowledges support from the European Research Council (ERC) through the starting grant 850899 – NEQuM.
    D.R. acknowledges support from the  Ministry  of  Electronics $\&$  Information  Technology  (MeitY),  India  under  the  grant for  ``Centre  for  Excellence  in  Quantum  Technologies''  with Ref. No. 4(7)/2020-ITEA. 
    
    \bibliography{Bibliography}

\end{document}